\documentclass[prl,aps,twocolumn,showpacs]{revtex4}
\usepackage{epsfig}
\usepackage{amssymb}
\usepackage{amsmath}
\usepackage{graphicx}
\usepackage{amsfonts}

\begin{document}

\title{Metal--Insulator Transition in 2D:
Experimental Test of the Two-Parameter Scaling}
\author{D.~A.~Knyazev$^{1}$, O.~E.~Omel'yanovskii$^{1}$,
V.~M.~Pudalov$^{1}$, I.~S.~Burmistrov$^{2,3}$}
\affiliation{$^{1}$P.~N.~Lebedev Physical Institute RAS, 119991
Moscow, Russia} \affiliation{$^{2}$L.D. Landau Institute for
Theoretical Physics RAS, Kosygina street 2, 119334 Moscow, Russia}
\affiliation{$^{3}$Department of Theoretical Physics, Moscow
Institute of Physics and Technology, 141700 Moscow, Russia}

\begin{abstract}
We report a detailed scaling analysis  of resistivity $\rho(T,n)$
measured for several high-mobility 2D electron systems in the
vicinity of the 2D metal-insulator transition. We analyzed the
data using the two parameter scaling approach and general scaling
ideas. This enables us to determine the critical electron density,
two critical indices, and temperature dependence for the
separatrix in the self-consistent manner. In addition, we
reconstruct the empirical scaling function describing a
two-parameter surface which fits well the $\rho(T,n)$ data.
\end{abstract}


\date{\today}

\maketitle

The experimental discovery~\cite{Pudalov1,prb95,Review1} of the
metal-insulator transition (MIT) in the two-dimensional (2D)
electron system of high mobility silicon metal-oxide-semiconductor
(Si-MOS) field-effect transistors still  calls  for a theoretical
explanation. Perhaps, the most promising framework is provided by
the microscopic theory, initially developed by Finkelstein, which
combines disorder and strong electron--electron (\textit{e-e})
interaction~\cite{Finkelstein}. Recently, Punnoose and
Finkelstein~\cite{PF_Science} have shown the possibility of the
MIT for a special model of the 2D electron system with the
infinite number of valleys. The key feature of the theory is the
existence of the \emph{two-parameter scaling} near the
criticality. One scaling variable is governed by disorder, and the
other is determined by \textit{e-e} interaction. In accord with
the theoretical prediction, the renormalization of both
resistivity and \textit{e-e} interaction with temperature has been
experimentally demonstrated \cite{KravchenkoNew,ParallelField}.

The main objective of the present Letter is to test experimentally
the two-parameter scaling. We present our results for the detailed
analysis of the temperature ($T$) and electron density ($n$)
dependence of resistivity ($\rho$) measured on five Si-MOS
samples. From this analysis, we identify the two scaling variables
together with the corresponding exponents (see
Table~\ref{samples}) and determine for the first time  the scaling
function for resistivity in the metallic region  in the vicinity
of the MIT. We have found that the two-parameter scaling provides
a natural explanation for the following set of experimental
observations in the vicinity of the MIT: (i) the separatrix,
$\rho_c(T) \equiv \rho(T,n_c)$, has a monotonic power-law
temperature dependence, where $n_c$ denotes the critical density;
(ii) a  generic $\rho(T,n)$ curve on the metallic side ($n>n_c$)
is nonmonotonic in $T$, and has a maximum and an inflection point;
(iii) for $n= n_c$, the maximum and inflection points  merge at
zero temperature; (iv) the normalized $\rho(T,n)/\rho_c(T)$ data
in the vicinity of the MIT  obey the mirror reflection symmetry
that is fulfilled in much wider ranges of $n-n_c$ and $T$, as
compared to the symmetry  previously reported under the assumption
of the $T$-independent separatrix
 \cite{Simonian97}.

\begin{figure}[h]
\includegraphics[width=190pt,height=260pt]{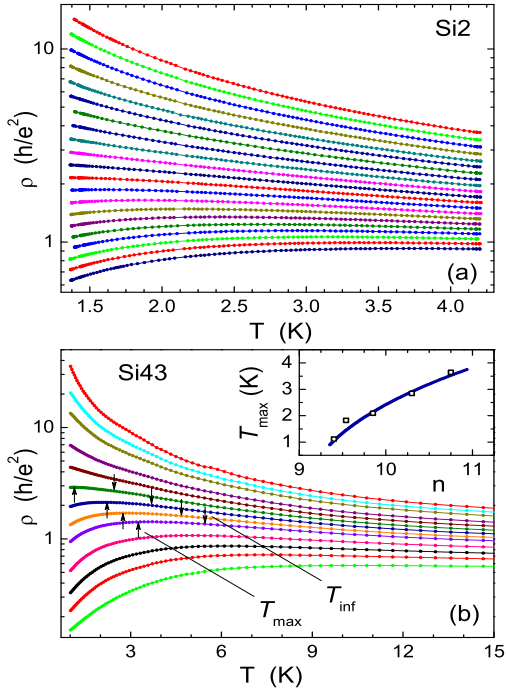}
\caption{Temperature dependences $\rho(T)$: (a) for sample Si2;
the densities increase in steps of 0.224 starting from 6.72
(topmost curve).  (b) for sample Si43; the densities  are (from
top to bottom) 7.16, 7.34, 7.52, 7.70, 7.88, 8.06, 8.51, 8.96,
9.41, 9.86, 10.31, 10.76, 11.6, 12.5, 13.4, 14.9. The inset shows
$T_{\rm max}$ vs electron density: squares are the data, and line
is the theoretical curve (see below). Density values are quoted in
units of $10^{10}$\,cm$^{-2}$.} \label{FigRawData}
\end{figure}

For the measurements we have selected five representative Si-MOS
samples of the rectangular geometry from four different (001)-Si
wafers. Their peak mobilities are listed in Table~\ref{samples}.
We have used four-terminal ac-technique at the 5\,Hz frequency.
The source--drain current was chosen low enough, e.g., $I=10$\,nA
for $T=1.3$\,K, in order to avoid electron overheating. Most of
 studies were performed in the range
$1.3-4.2$\,K, because for these temperatures (i) the critical
region of densities $|\Delta n|=|n-n_c|$ becomes sufficiently wide
as will be shown below, and (ii) $\rho(T)$ near the criticality
exhibits a well-pronounced maximum (Fig.~\ref{FigRawData}), which
provides an additional tool for comparison with
the theory. For two samples, Si15 and Si62, we have performed the
measurements down to 0.3\,K; for sample Si43, we have extended
the temperature range up to $1-38$\,K in order to illuminate the
overall trend of the resistivity maximum and the scaling behavior
of $\rho(T,n)$.

Figure~\ref{FigRawData} shows a typical $\rho(T)$-dependence  for
two samples, Si2 (Si43), in the critical regime,
in the density range from 0.672 to 1.12 (0.707 to 1.49), in units of
$10^{11}$\,cm$^{-2}$ \cite{density}. As $n$ decreases,
the $\rho(T)$ behavior crosses
over from metallic ($n>n_c$) to insulating ($n<n_c$) one.
Initially and during about a decade \cite{prb95,Review1,Simonian97}, the data were
analyzed within the framework of the one-parameter scaling (OPS)
theory,  in which $n$ is a single driving parameter.
Within this approach the separatrix $\rho_c(T)$
separating the metallic from insulating domains
should be $T$-independent, i.e, $\rho_c(T)=\rho_0$. As a result,
in the vicinity of $n_c$, the $\rho(T,n)$ data should obey the
reflection symmetry \cite{dobro97}: $\rho(T,n_c-\Delta n) =
\rho_0^2/\rho(T,n_c+\Delta n)$ when $\Delta n \ll n_c$.
This obviously disagrees with  the data shown in
Fig.~\ref{FigRawData}, i.e., the OPS theory cannot adequately describe
the data within the whole $T$-range. In view
of this well-known problem \cite{noscaling}, it is a common
practice to  use the OPS approach for the
analysis of the data in a  \emph{truncated} temperature range.

In theory~\cite{Finkelstein}, physics at $T=0$ is described by the
two coupled renormalization group (RG) equations for conductance
$\sigma$ (in units $e^2/h$) and interaction amplitude $\gamma_2 =
-F_0^\sigma/(1+F_0^\sigma)$ where $F_0^\sigma$ denotes the
standard Fermi-liquid parameter in the triplet
channel~\cite{Finkelstein,Italians}:
\begin{equation}
\frac{d\sigma}{d\eta} = \beta_\sigma(\sigma,\gamma_2),\qquad
\frac{d\gamma_2}{d\eta} =
\beta_{\gamma_2}(\sigma,\gamma_2).\label{Eq:1}
\end{equation}
Here $\eta=\ln L/l$, $L$ stands for the running RG lengthscale
(sample size),
 and $l$ is the microscopic lengthscale (elastic mean free path),
  at which the RG starts. In the microscopic
theory~\cite{Finkelstein}, the temperature appears as $T
z$ which has a dimensionality of $L^{-2}$, and the parameter $z$
varies with
 $L$ according to $d\ln z/d\eta  =
\gamma_z(\sigma,\gamma_2)$. The RG equations should be
supplemented with the initial conditions at $\eta=0$: $\sigma(0) =
\bar\sigma$, $\gamma_2(0)=\bar\gamma_2$, and $z(0) = \bar z$,
where $\bar\sigma, \bar\gamma_2$ and $\bar z$ are sample
dependent. Following the scaling ideas~\cite{Amit},
we expect the measured resistivity $\rho = \rho(T\bar
z,\bar\sigma,\bar\gamma_2,l)$
to be independent of microscopic details: $d\rho/d l =0$.
Applying the standard analysis of the Callan--Symanzyk
equation~\cite{Amit}, we find that the measured resistance is a
function of two scaling variables $X$ and $Y$,
$\rho=\mathcal{R}(X,Y)$, where $\mathcal{R}$ is a regular
function.

The MIT implies the existence of a fixed point (FP) determined by
the conditions: $\beta_{\sigma}(\sigma^c,
\gamma_2^c)=\beta_{\gamma_2}(\sigma^c, \gamma_2^c)=0$. Linearizing
Eqs.~\eqref{Eq:1} near the fixed point, one finds
\begin{equation}
X = (T/T_0)^{-\kappa} (n-n_c)/n_c,\quad Y = (T/T_1)^{\zeta}.
\label{Eq:2}
\end{equation}
Here $\kappa = p/(2\nu)$ and $\zeta= -p y/2$, where the
correlation length exponent $\nu$ and the irrelevant exponent
$y<0$ are given by the eigenvalues of the linearized
Eqs.~\eqref{Eq:1}~\cite{Amit}. The exponent $p=
2/[2+\gamma_z(\sigma^c,\gamma_2^c)]$ governs the $T$-behavior of
the specific heat at the fixed point;  $T_{0,1}$ are
sample-dependent energy scales.  Temperature $T_0$ corresponds to
the bandwidth of the available states participating in the
transport and is of the order of the elastic scattering rate
$\hbar/2k_B\tau$.  Temperature $T_1$ determines the quantum
critical region at $n=n_c$ \cite{Amsterdam}.  Deriving
Eq.~\eqref{Eq:2}, we expressed $\bar\sigma$ and $\bar\gamma_2$ in
terms of the experimentally measurable parameters $\mu$ and $n$.
\begin{figure}[h]
\includegraphics[width=200pt,height=260pt]{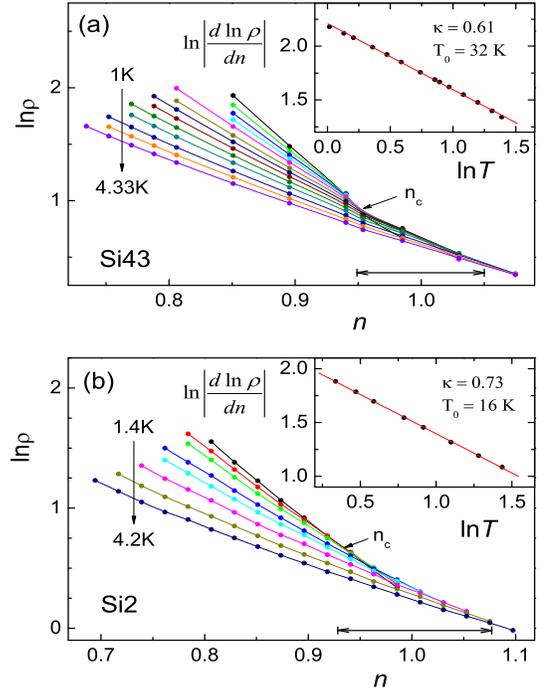}
\caption{The density dependences of $\ln\rho(n)$ for various
temperatures (from the top to the bottom, in K): (a) $1, 1.23,
1.43, 1.6, 1.8, 2.1, 2.35, 2.64, 2.95, 3.31, 3.71, 4, 4.33$ for
sample Si43, (b) $1.4, 1.6, 1.8, 2.2, 2.5, 3, 3.6, 4.2$ for sample
Si2. The density is given in units of $10^{11}$\,cm$^{-2}$. The
data are shown in a limited density range, $X<0.7$. The insets
illustrate the procedure of extracting $\kappa$ and $T_0$ using
Eq.~\protect\eqref{Eq:3}. The bars display the range of
intersections for all curves.}\label{FigKappaPlus}
\end{figure}

At present, the explicit form of $\mathcal{R}(X,Y)$ is unknown. In
general, even in the  vicinity of the FP, $X$ and $Y$ are not
necessarily small. However, from the common-sense arguments we
require that at $T\to 0$ the system (i) is  an insulator for
$n<n_c$, i.e., $\mathcal{R}(-\infty,0) = \infty$; (ii) is a metal
for $n>n_c$, i.e., $\mathcal{R}(+\infty,0) = 0$;  and (iii) is  at
the quantum critical point for $n=n_c$, i.e., $\mathcal{R}(0,0) =
\rho_c^0$. Following arguments of Ref.~\cite{dobro97}, we assume that
\begin{equation}
\rho(T,n) = \mathcal{R}(X,Y)= \rho_c^0 e^{-X} (1-Y), \quad |X|,\,
Y\ll 1. \label{Eq:3}
\end{equation}
This is consistent with all the requirements listed above. In
particular, such a form of $\mathcal{R}(X,Y)$ provides the
reflection symmetry for the normalized resistance
$\rho(T)/\rho_c(T)$.

In the vicinity of $n_c$, on the metallic side, $n>n_c$, the
$\rho(T,n)$ data exhibit a maximum at a certain density-dependent
temperature $T_{\rm max}$ (see Fig.~\ref{FigRawData}). The inset
to Fig.~\ref{FigRawData} shows that $T_\textrm{max}$ vanishes as
$n \to n_c$, in agreement with Eq.~\eqref{Eq:3}, which predicts a
monotonic $\rho_c(T)$ dependence. On the insulating side, for $n$
lower but close to $n_c$, the $\rho(T,n)$ curves have a positive
curvature. Therefore, the neighboring curves on the metallic side
should also have a positive curvature at sufficiently high
temperatures. This expectation agrees with the experiment
(Fig.~\ref{FigRawData}).
It follows from the above results that each of the nonmonotonic
$\rho(T,n)$ curves must have \emph{an inflection point}. Its
existence restricts the exponent $\zeta$ to be less or equal to
unity, $\zeta\leqslant 1$. As a result, not only $T_\textrm{max}$
but also  the inflection point temperature $T_{\rm inf}$ vanish at
$n=n_c$.

In order to test the theoretical prediction, we have adopted a
procedure employed for studying the scaling behavior
near the criticality in the quantum Hall regime~\cite{Amsterdam}.
We plot in Fig.~\ref{FigKappaPlus} the $\ln \rho(T,n)$ data for
various temperatures versus $n$ for $X\lesssim 1$. Two features
are clearly seen in Fig.~\ref{FigKappaPlus}. Firstly, as $T$
decreases, the region of intersection for the $\ln\rho(T,n)$
curves shrinks. This enables one to determine an approximate $n_c$
value as an intersection point in the $T=0$ limit. Secondly, in
the vicinity of $n_c$  (for small $X$), the $\ln\rho(T,n)$ curves
are linear in $n$ in accordance with Eq.~\eqref{Eq:3}. In the
insets to Fig.~\ref{FigKappaPlus}, we plot the logarithmic slopes
of the $\ln\rho(T,n)$ curves versus $\ln T$. The resulting curves
appear to be linear; this allows us to extract both $\kappa$ and
$T_0$ values for the five studied samples. The results are
summarized in Table~\ref{samples}.

\begin{figure}[h]
\includegraphics[width=190pt]{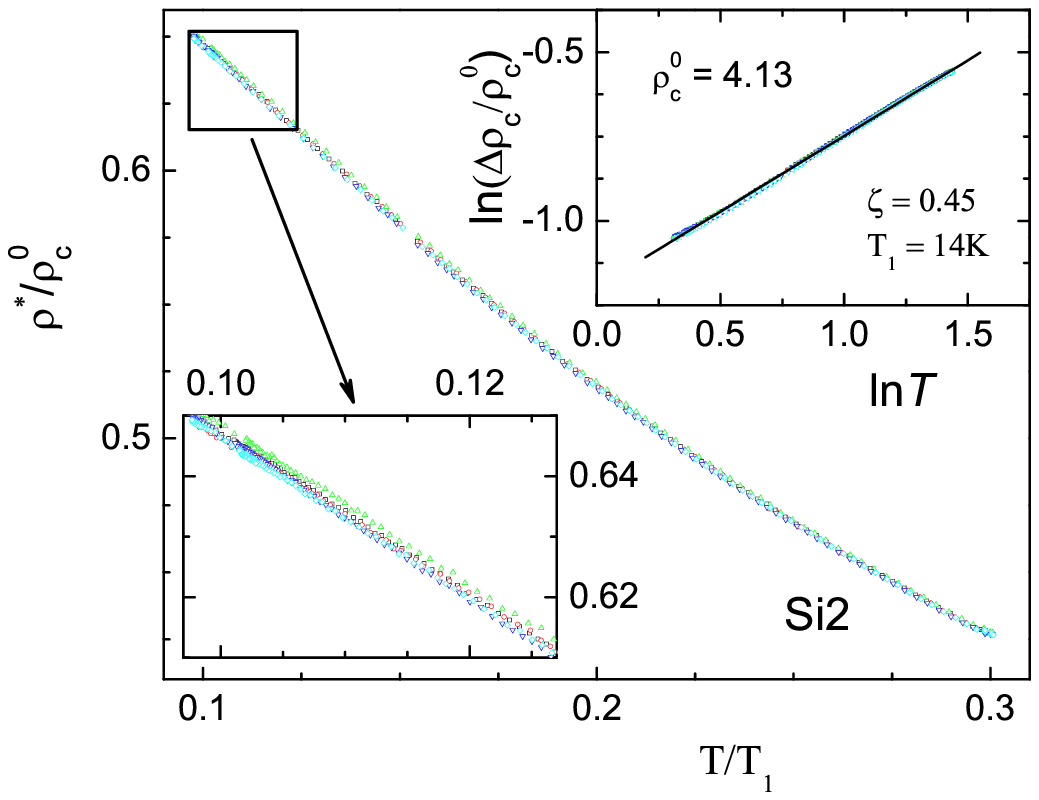}
\caption{Scaling of the $\rho^*(T)/\rho_c^0$ data for densities
$n=8.74, 8.96, 9.18, 9.41, 9.63 \times 10^{10}$\,cm$^{-2}$. The lower
inset demonstrates the quality of the scaling for large $X \sim
0.5$. The upper inset shows a fit of the separatrix using
Eq.~\eqref{Eq:3} with three parameters $\zeta$, $T_1$ and
$\rho_c^0$. Sample Si2}\label{FigKappaMinus}
\end{figure}

Equation~\eqref{Eq:3} predicts that, for $X\ll 1$, the $\rho(T,n)$
data normalized by $\exp(-X)$ should
collapse onto a single curve. In order to test this theoretical
prediction, we plot the normalized $\rho^*(T) \equiv \rho(T,n)
\exp(X)$ data in Fig.~\ref{FigKappaMinus}. We stress that no
adjustable parameters are used in this scaling procedure. The
critical density $n_c$,
and the $T_0$ and $\kappa$
values were determined at the previous stage. As is shown in
Fig.~\ref{FigKappaMinus}, the normalized $\rho^\star(T)$ data
 do scale with high accuracy, 0.8\%, on the $\rho(T,n_c)$
curve in the range $X<0.5$. The deterioration  of the scaling
quality for
greater $X$ values limits the range of temperatures and densities
used for the scaling procedure (Fig.~\ref{FigKappaMinus}).
Provided that Eq.~\eqref{Eq:3} adequately describes  the
separatrix $\rho(T,n_c)$, the scaling procedure enables us to
determine the critical $n_c$ value more precisely. For this
purpose, we require the low temperature region ($Y \ll 1$)  of the
scaled curves to have the curvature of the constant sign over the
whole temperature range. Indeed, we  found that, by adjusting the
preliminary determined $n_c$ value within  only 2\%, the
inflection point in the $\rho(T,n_c)$-data could be easily
eliminated for all five samples.

The bunch of the scaled curves in Fig.~\ref{FigKappaMinus} is
fitted with the theoretical expression~\eqref{Eq:3} for the
separatrix $\rho(T,n_c)$, by using the remaining scaling
parameters $\zeta$, $T_1$ and $\rho_c^0$ as adjustable parameters.
The result of fitting for sample Si2 is shown in the inset to
Fig.~\ref{FigKappaMinus}. The
values of the above parameters extracted from the best fit are
given in Table~\ref{samples} for all of the five samples.
The validity of Eq.~\eqref{Eq:3} is further supported by the
nonmonotonic temperature behavior of $\rho(T,n)$
(Fig.~\ref{FigRawData}). Using the extracted scaling parameters
(Table~\ref{samples}), we can readily find $T_{\rm max}(n)$
dependences from Eq.~\eqref{Eq:3}. The results are in a good
agreement with the experimental data for all of the five samples.
A typical example of comparison is shown in the inset to
Fig.~\ref{FigRawData}. We note that the data are processed only
for $T\leq (4-5)$\,K to ensure the applicability of
Eq.~\eqref{Eq:3}.
\begin{table}[h]
\caption{The relevant parameters for the studied samples.
$\mu_{peak}$ [m$^2$/Vs] is the peak mobility at $T =0.3$\,K,
$\rho_c^0$ is in units of $h/e^2$, $T_{0}$ and $T_1$ are in K, and
$n_c$ is in $10^{11}$\,cm$^{-2}$. The inverse transport scattering
time $\hbar/2k_B\tau$ [$K$] is estimated from $\rho_c(T=4.2\,K)$.
The errors for $T_0$, $\kappa$, $n_c$ are about 2\%;  the errors
for $T_1$, $\zeta$, and $\rho_c^0$
 are given in the last column.}
\begin{tabular}{|c||c|c|c|c|c|c|c|c|c|}
\hline sample & $\mu_{\rm peak}$ & $\hbar/2k_B\tau$ & $n_c$ & $T_0$ &
$T_1$ & $\kappa$ & $\zeta$& $\rho_c^0$ & \%\\
\hline Si15  & 4.1 & 13 & 0.86 & 12 & 15   & 0.82  & 0.35 & 4.17 & 30\\
Si62   & 3.6 & 26& 0.94 & 22 & 11 & 0.74  & 0.5 & 5.27 & 20\\
Si2    &3.4 &22& 0.89 & 16 & 14 & 0.73 & 0.45 & 4.13 &12\\
Si43   &2.0 &31& 0.92 & 32 & 13.5 &  0.61 & $0.8$&
4.14 &12 \\
Si6-14 & 1.9 &26&  1.22 & 26 & 15   &  0.66 & 0.86& 2.35 & 6\\
\hline
\end{tabular}
\label{samples}
\end{table}

Table~\ref{samples} shows that the extracted  $T_0$ and $T_1$
agrees reasonably with $\hbar/2k_B\tau$. In the theory, the
exponents $\kappa$, and $\zeta$, and the $\rho_c^0$ value should
be universal. Experimentally, there is a trend of $\kappa$ and
$\zeta^{-1}$ towards smaller values with decreasing sample
mobility. The trend is rather likely to be due to the presence of
the long-range components of the random potential for the high
mobility Si-MOS samples. This favors a percolation-type behavior
to dominate over the true scaling one \cite{baranov}.
We mention that variation of exponent
$\kappa$ with a sample  mobility has recently been observed in the
plateau--plateau transitions of the quantum Hall regime~\cite{Li}.
Yet another possible reason that may complicate the scaling
analysis of experimental data is an inhomogeneity of the electron
density~\cite{Inhom}.

In contrast to the one-parameter scaling, in the two-parameter
case the data should be described by a universal scaling function
$\mathcal{R}(X,Y)$, which  represents a 2D surface in the
$(X,Y,\rho)$ space. Using six parameters determined
experimentally, $n_c, \kappa, \zeta, T_{0,1}$ and $\rho_c^0$
(Table 1), we calculate the $(\rho/\rho_c^0)$ data versus $X$ and
$Y$. The data (64000 points) for all samples collapse close to a
single surface in the $(X,Y,\rho/\rho_c^0)$ space. To model the
surface analytically, we use $\mathcal{R}(X,Y)=
\rho_c^0\exp[f_1(X)] f_2(Y)$ and fit the functions $f_{1,2}$  with
five adjustable parameters as $f_1= -X + 0.07 X^2 + 0.001 X^3$ and
$f_2 = (1-Y+ 1.48Y^2)/(1+1.88 Y^2 +1.72Y^3)$. This procedure
yields an agreement of the model surface with the data  within 4\%
rms in the range $|X| \leq 5$ and $Y \leq 3$. The $X^2$ term in
$f_1$ describes the violation of the reflection symmetry for large
$X$. The analysis of our $\rho(T)/\rho_c(T)$  data confirms that
the reflection symmetry $X \leftrightarrow -X$ holds
within 0.8\% accuracy in the region $|X|< 0.5$, $Y<0.7$ (see
Fig.~3).
\begin{figure}[h]
\includegraphics[width=220pt]{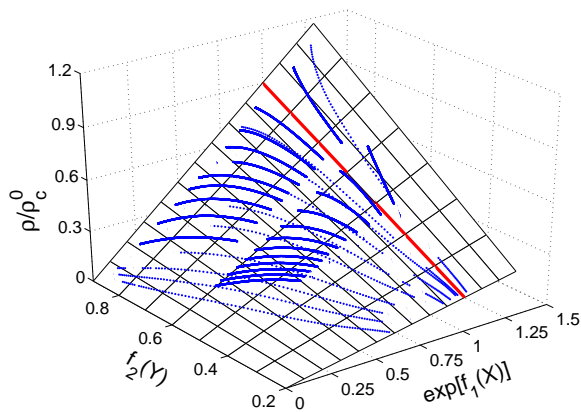}
 \caption{The empirical
two-parameter scaling function  $\mathcal{R}(X,Y)$
 for five samples (dots) and the model surface
 $\mathcal{F} = \tilde{X}\tilde{Y}$  (grid).
The straight line $\rho/\rho_c^0 = \tilde{Y}$ at $\tilde{X} = 0$
represents the separatrix.} \label{FigScalingFunction}
\end{figure}

The resulting
model surface $\mathcal{F}=\tilde{X}\tilde{Y}$ is plotted in
Fig.~\ref{FigScalingFunction} as a function of $\tilde{X} =
\exp[f_1(X)]$ and $\tilde{Y}= f_2(Y)$ together with the scaled
data $\rho(T,n)/\rho_c^0$. The data are very close to the model
surface; the maximal deviation of the data from the model surface
in the $\rho$ direction does not exceed $0.05\rho_c^0$. Clearly,
the surface is curved in the both directions. The separatrix
$\rho_c(T)$ is represented by the straight line on the model
surface (Fig.~\ref{FigScalingFunction}).

In summary, we have performed  a detailed two-parameter scaling
analysis of the $\rho(T,n)$ data for five Si-MOS samples in the
vicinity of the 2D metal-insulator transition. The two-parameter
scaling function, which we have determined experimentally,
naturally incorporates the temperature dependence of the
separatrix $\rho_c(T)$, the generalized mirror reflection symmetry
with respect to $\rho_c(T)$, and the existence of the maximum and
inflection points for the $\rho(T)$ dependence on the metallic
side. Our analysis strongly supports the interpretation of the
critical behavior observed in the transport data as the
manifestation of the quantum phase transition driven by both
disorder and interaction. The sample to sample variations observed
in the exponents $\kappa$ and $\zeta$, and
in the critical resistivity $\rho_c^0$ value require a quest for
new samples in which a random potential will have much shorter
correlation length and the inhomogeneities
of the electron density  will be reduced.

The authors are grateful to A.M.M.~Pruisken and A.~M.~Finkelstein for discussions.
The
work was supported in part by the RFBR, Programs of RAS, Russian
Ministry for Education and Science, Program ``The State Support of the
Leading Scientific Schools'', Russian Science Support Foundation
(D.A.K.), CRDF, FASI, and Dynasty Foundation (I.S.B.).


\begin{thebibliography}{99}
\bibitem{density}The absolute values for the carrier density
specified in the paper are determined from the period of the
quantum magnetooscillations within 1\% error, including the
vicinity of $n_c$~\protect\cite{granada}. The relative deviation
in the density $\Delta n/n_c$ is controlled much more precisely,
within $0.1\%$, from the reproducibility of the $\rho(n)$ data for
a fixed $T$.
\end{thebibliography}
\end{document}